\documentclass[
]{ceurart}

\usepackage{listings}
\begin{document}

\copyrightyear{2024}
\copyrightclause{Copyright for this paper by its authors.
  Use permitted under Creative Commons License Attribution 4.0
  International (CC BY 4.0).}

\conference{CHIRP 2024: Transforming HCI Research in the Philippines Workshop, May 09, 2024, Binan, Laguna}

\title{Simon Says: Exploring the Importance of Notification Design Formats on User Engagement}


\author[]{Hans Matthew {Abello}}[%
email=hans_abello@dlsu.edu.ph,%
]
\author[]{Maxine Beatriz {Badiola}}[%
email=maxine_badiola@dlsu.edu.ph,%
]
\author[]{Mark John {Custer}}[%
email=mark_john_custer@dlsu.edu.ph,%
]
\author[]{Lorane Bernadeth {Fausto}}[%
email=lorane_bernadeth_fausto@dlsu.edu.ph,%
]
\author[]{Patrick Josh {Leonida}}[%
email=patrick_josh_c_leonida@dlsu.edu.ph,%
]
\author[]{Denzel Bryan {Yongco}}[%
email=denzel_bryan_yongco@dlsu.edu.ph,%
]
\author[]{Jordan Aiko {Deja}}[%
email=jordan.deja@dlsu.edu.ph,%
orcid=0001-9341-6088%
]

\address[1]{De La Salle University, Manila, Philippines}

\begin{abstract}
Push notifications are brief messages that users frequently encounter in their daily lives. However, the volume of notifications can lead to information overload, making it challenging for users to engage effectively. This study investigates how notification behavior and color influence user interaction and perception. To explore this, we developed an app prototype that tracks user interactions with notifications, categorizing them as accepted, dismissed, or ignored. After each interaction, users were asked to complete a survey regarding their perception of the notifications. The study focused on how different notification colors might affect the likelihood of acceptance and perceived importance. The results reveal that certain colors were more likely to be accepted and were perceived as more important compared to others, suggesting that both color and behavior play significant roles in shaping user engagement with notifications.
\end{abstract}


\maketitle

\section{Introduction and Background}
\par Push notifications have become an integral part of modern communication, reshaping how we receive messages across various contexts, including social, work, and emergency notifications. These notifications deliver concise, relevant information in a short amount of time, enabling users to quickly assess whether they should engage further or ignore them.

\par Apple has introduced Focus Mode to experiment with notification behavior, building on its ``Do Not Disturb Mode'' \cite{evans_2021}. Focus Mode allows users to tailor notification delivery by limiting the types of notifications they receive and assigning them varying levels of priority. In contrast, Android users benefit from more customization options, including a snooze feature that lets users temporarily silence notifications while ensuring they do not miss important messages \cite{gilbert_how_2024}.

\par In today’s digital landscape, notifications are essential for keeping users connected to both their virtual and real-world obligations. However, with the increasing volume of notifications, managing them efficiently has become a critical design challenge. A study by \citeauthor{pielot_2014} \citeyear{pielot_2014} found that users receive an average of 63.5 notifications per day, and many report feelings of stress or overwhelm due to the constant influx of alerts.

\par To address this challenge, streamlining push notifications for easier assessment, readability, and accessibility is crucial. Given the customization options available, it is important to systematically understand how users interact with notifications, considering factors such as content and the time it takes to respond. Additionally, this study investigates how notification color influences the likelihood of user interaction or dismissal. Previous research has shown that colors can significantly affect cognitive, psychological, and emotional responses, influencing user behavior \cite{elliot_color_2007, achint_kaur_link_2020, valdez_effects_1994, silic_2016}. In this study, notifications are characterized by two variables: color and category. Three colors and three categories were tested, each with its own heading and body text, as detailed in Table \ref{tab:catdes}.

\begin{table}[h]
\caption{Category descriptions}
\label{tab:catdes}
\begin{tabular}{|p{2cm}|p{2cm}|p{0.75cm}|p{6.5cm}|}
\hline
\textbf{Category} & \textbf{Header} & \textbf{Color} & \textbf{Body}                                                                                 \\ \hline
Emergency                          & Earthquake Alert: Seek Shelter   & red   & \texttt{NDRRMC: (1:25PM,13Mar24) Isang Magnitude 7.3 na lindol ang naganap kaninang 1:05PM.  Aftershocks ay inaasahan.} \\ \hline
Social                             & Facebook Messenger               & green & \texttt{John Doe sent you a friend request. }                                                                           \\ \hline
Work                               & Canvas Student                   & blue  & \texttt{New Assignment: \{{[}\}Problem Set\{{]}\} Cache Memory Analysis: \{{[}\}1232\_CSARCH2\_S11\{{]}\} }             \\ \hline
\end{tabular}
\end{table}

\par Notifications also allow for user interaction, with three primary actions: accepting, dismissing, or ignoring. \textbf{\textit{Accept}} indicates that the user intends to view the content of the notification. \textbf{\textit{Dismiss}} means the user has decided not to engage with the notification content. \textbf{\textit{Ignore}} is used when a notification is left unattended for at least 10 seconds, marking it as ignored. Additionally, users can configure notification behaviors, which can influence the perceived urgency of the notifications, as outlined in Table \ref{tab:notifdes}.

\begin{table}[h]
\caption{Notification behaviors descriptions}
\label{tab:notifdes}
\begin{tabular}{|l|p{6.5cm}|}
\hline
\textbf{Behavior} & \textbf{Description}                                                                                                                         \\ \hline
Urgent            & Notifications that were ignored would disappear.                                                                                             \\ \hline
Non-urgent        & Notifications that were ignored would not disappear, allowing the user to still interact with them. These notifications can stack with one another. \\ \hline
\end{tabular}
\end{table}

\par Interaction time, denoted as $IT$, refers to the duration it takes for a user to interact with a notification, measured in seconds. This is calculated by subtracting the time the notification was shown ($T_s$) from the time the notification was interacted with ($T_r$), as expressed in the following equation:

\begin{equation} IT = T_r - T_s \end{equation}

Where $T_r$ is the time when the notification is interacted with, and $T_s$ is the time when the notification first appears.

\par The primary contribution of this paper lies in exploring how notification colors and their format (e.g., content and urgency) influence user engagement. By examining how different color schemes and notification characteristics affect user interaction time and behavior, this study aims to provide insights into optimizing notifications for better user experience and engagement.



\section{Method}
\subsection {Protocol}
\par An online prototype was developed in the form of a game inspired by ``Simon Says,''~\cite{strommen1973verbal} featuring a wheel with four colors. The game begins with a randomly flashing color on the wheel, and the user’s objective is to remember and reproduce the randomly generated sequence of colors. Each time the user correctly selects the right color, additional colors are added to the sequence, thereby increasing the user’s score. The game ends once the user reaches a score of 8. However, if the user makes an error, the sequence restarts. During gameplay, notifications were displayed at random score intervals in the upper-right corner of the screen.

\par Ten university students participated in this experiment. Each participant completed the experiment twice, once with urgent notifications and once with non-urgent notifications. The experiment began once the user selected a notification behavior, and it concluded when they had viewed all possible notification combinations. The results were automatically generated at the end of the experiment.

\subsection{Post-Experiment Analysis}
\par Upon completing the experiment, participants were asked to fill out a post-experiment survey. The survey included questions with a 5-point Likert scale, categorical questions, and open-ended questions for qualitative responses. The objective of the survey was to gather participants’ perceptions of the importance of each notification, their color preferences for the notifications, and their overall feedback regarding the experiment.

\section{Results}
\par A total of 180 data points were collected for each behavior in the experiment. The rows where the status was labeled as 'ignored' were excluded from the dataset, as these observations did not provide interaction times. This resulted in 136 entries for the urgent behavior and 144 entries for the non-urgent behavior.


\subsection{How does interaction time vary across different colors and statuses in each dataset?}

\begin{table}[h]
    \caption{Descriptive statistics for color in both behaviors}
    \label{tab:urgent-color}
    \begin{tabular}{|l||l||l|l|l|l|l|}
        \hline
         \textbf{behavior}&\textbf{color} & \textbf{count} & \textbf{mean} & 
            \textbf{std} & \textbf{min} & \textbf{max}\\ \hline
         Urgent&Blue & 46.0 & 2.310& 1.679& 0.704 & 8.280 \\ \hline
         Urgent&Green & 43.0 & 1.847& 1.281& 0.821 & 5.924 \\ \hline
         Urgent&Red & 47.0 & 2.225& 1.768& 0.689 & 6.741 \\ \hline
 Non-urgent& Blue 
& 48.0 & 2.780& 3.70& 0.694 &18.692 
\\\hline
 Non-urgent& Green 
& 48.0 & 6.315& 13.579& 0.641 &67.916 
\\\hline
 Non-urgent& Red & 48.0 & 3.357& 8.152& 0.754 &53.551 \\\hline
    \end{tabular}
\end{table}

\par For urgent notifications, as shown in Table \ref{tab:urgent-color}, the color \textit{blue} has a mean interaction time of 2.310 seconds, with times ranging from 0.704s to 8.280s. The color \textit{green} has a mean time of 1.847s, with a range from 0.821s to 5.924s. The color \textit{red} has a mean time of 2.225s, with a range from 0.689s to 6.741s. Among these colors, \textit{blue} has the highest mean interaction time, while \textit{green} has the lowest.

\par In contrast, for non-urgent notifications, as shown in Table \ref{tab:urgent-color}, the color \textit{blue} has a mean time of 2.781s, with a range from 0.694s to 18.692s. The color \textit{green} has a mean time of 6.315s, ranging from 0.641s to 67.916s. The color \textit{red} has a mean time of 2.225s, with a range from 0.689s to 6.741s. Among these colors, \textit{green} has the highest mean interaction time, while \textit{blue} has the lowest.



\subsection{How does interaction time vary across different categories and statuses in each dataset?}

\begin{table}[h]
    \caption{Descriptive statistics for category in urgent behavior}
    \label{tab:urgent}
    \begin{tabular}{|l||l||l|l|l|l|l|}
        \hline
         \textbf{behavior}
&\textbf{category} & \textbf{count} & \textbf{mean} & 
            \textbf{std} & \textbf{min} & \textbf{max}\\ \hline
         Urgent
&Emergency & 44.0 & 2.146& 1.543& 0.689 & 6.741 \\ \hline
         Urgent
&Social & 47.0 & 2.263& 1.821& 0.747 & 8.280 \\ \hline
         Urgent
&Work & 45.0 & 1.987& 1.414& 0.694 & 6.582 \\ \hline
  Non-urgent
&Emergency & 48.0 & 4.778& 12.179& 0.641 &67.916 
\\\hline
  Non-urgent
&Social & 48.0 & 2.756& 5.387& 0.787 &33.455 
\\\hline
  Non-urgent&Work & 48.0 & 4.920& 9.569& 0.694 &53.551 \\\hline
    \end{tabular}
\end{table}

\par For urgent notifications, as shown in Table \ref{tab:urgent}, the \textit{Emergency} category had interaction times ranging from 0.689s to 6.74s, with a mean time of 2.146s. The \textit{Social} category had times ranging from 0.747s to 8.280s, with a mean time of 2.263s. The \textit{Work} category had times ranging from 0.694s to 6.582s, with a mean time of 1.987s. Among these categories, \textit{Work} had the smallest mean interaction time, while \textit{Social} had the largest.

\par In contrast, for non-urgent notifications, as shown in Table \ref{tab:urgent}, the \textit{Emergency} category had a mean interaction time of 4.778s, with times ranging from 0.641s to 67.916s. The \textit{Social} category had a mean time of 2.755s, with times ranging from 0.787s to 33.455s. The \textit{Work} category had times ranging from 0.694s to 53.551s, with a mean time of 4.919s. Contrary to the urgent dataset, the \textit{Social} category had the smallest mean interaction time for non-urgent behavior, while \textit{Work} had the largest.



\subsection{Is there any difference in the relationship between color and status between the two datasets?}





\begin{figure}[htbp]
    \centering
    \includegraphics[width=0.9\textwidth]{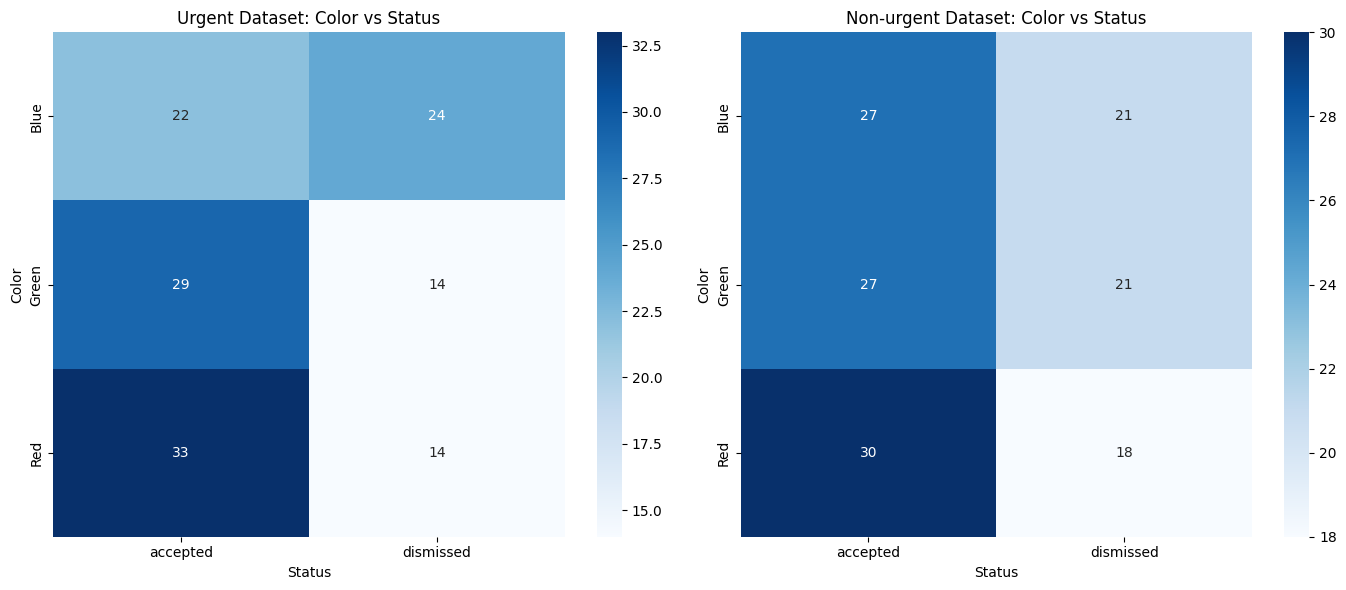}
    \caption{Color vs Status Heatmap}
    \label{fig:heatmap}
\end{figure}

\par The heatmaps represented in Figure \ref{fig:heatmap} show the number of occurrences of specific color-status combinations within the urgent dataset. The color intensity in each cell reflects the count of observations: darker shades indicate higher counts, while lighter shades represent lower counts.

\par In the urgent dataset, for the color Red, there are 33 accepted interactions, 14 dismissed interactions, and 3 ignored interactions. These are represented by darker shades in the respective cells. For the color Green, 29 interactions were accepted, 14 were dismissed, and 3 were ignored. Similarly, for the color Blue, there were 22 accepted interactions, 24 dismissed, and 3 ignored.

\par In the non-urgent dataset, for the color Red, there are 30 accepted interactions, 18 dismissed interactions, and 3 ignored interactions. For the color Green, 27 interactions were accepted, 21 were dismissed, and 3 were ignored. Finally, for the color Blue, 27 interactions were accepted, 21 were dismissed, and 3 were ignored.

\par Overall, the behavior shown by participants for the color Red was consistent across both scenarios, with Red having the highest number of accepted interactions in the dataset. Additionally, it is notable that there were generally more dismissed interactions in the non-urgent dataset, which is expected, as the level of urgency is lower compared to the urgent dataset.

\subsection{Are there any differences in the average interaction times between the urgent and non-urgent datasets?}

The mean times for urgent and non-urgent behaviors are \textbf{2.134s} and \textbf{4.151s}, respectively. Given the mean times for the two behaviors in Table \ref{tab:means}, the t-test statistic was used to determine the significance of the difference in both means. With $\alpha = 0.05$, the computed t-value and p-value is $-2.455$ and $0.014$, respectively. This states that there is a significant difference between the two means shown in Figure \ref{tab:means}

\subsection{Is there any correlation between category and status in each dataset?} 

\begin{figure}[htbp]
    \centering
    \includegraphics[width=0.9\textwidth]{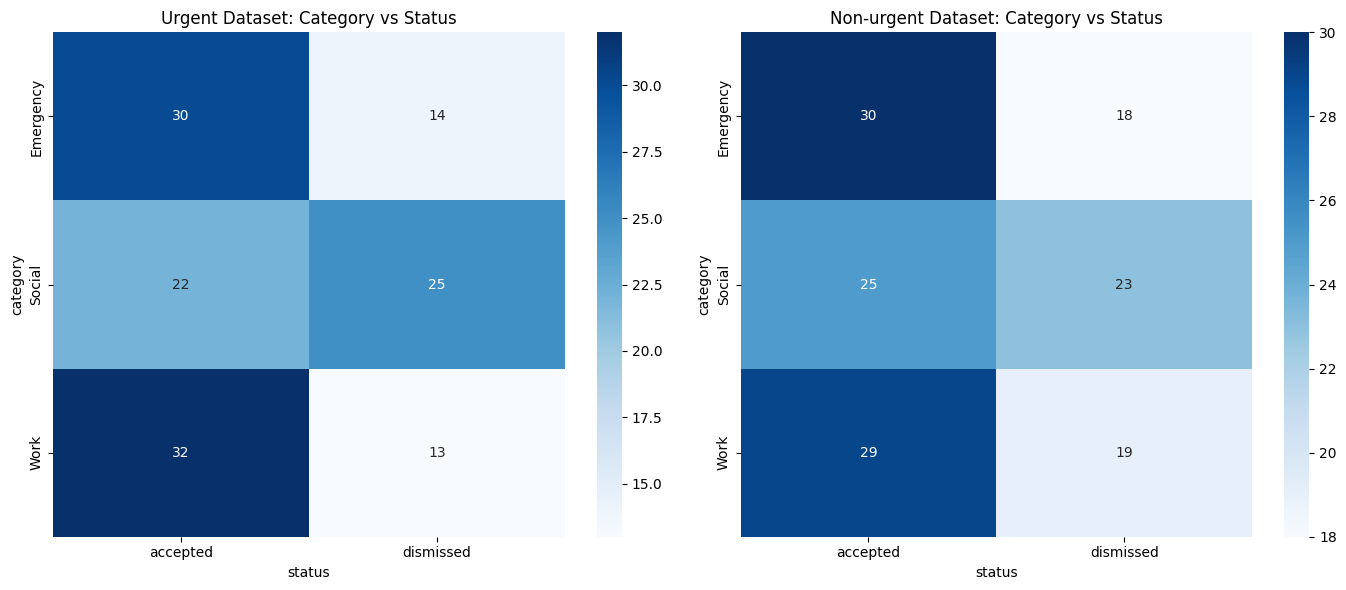}
    \caption{Category vs Status Heatmap}
    \label{fig:heatmap}
\end{figure}

\par The heatmaps display the contingency tables for both datasets, with rows representing categories (Emergency, Work, Social) and columns representing statuses (Accepted, Dismissed, Ignored). A chi-square test was applied to the contingency tables of both datasets to assess whether the two variables are independent of each other.

\par For the urgent dataset, the calculated p-value is approximately $0.03$. With $\alpha = 0.05$ and a chi-square statistic of $\chi^2 = 6.883$, this indicates that the variables are dependent on each other. In contrast, for the non-urgent dataset, the chi-square test yields a p-value of $0.548$ and a chi-square statistic of $\chi^2 = 1.200$, suggesting that the variables are independent.



\subsection{Do notification interactions differ between urgent and non-urgent contexts?}
\par After conducting our exploratory data analysis, we observed significant differences in interaction behavior between the urgent and non-urgent datasets.

\par The visual graphs display similar, if not identical, trends, suggesting that users interact differently depending on the level of urgency of a given notification. However, further analysis revealed that factors such as category, status, and interaction time do not vary significantly and do not exhibit a strong correlation in non-urgent notifications. In contrast, there is a significant correlation between category and status in urgent notifications.



\subsection{Participant Opinions}

\par After the tests, the participants were surveyed to assess their perceived importance of the different categories and colors. The first part of the survey used a 5-point Likert scale to measure the perceived importance of each category and color, as shown in Table \ref{tab:likert}.

\par
\begin{figure}[htbp]
    \centering
    \includegraphics[width=0.9\textwidth]{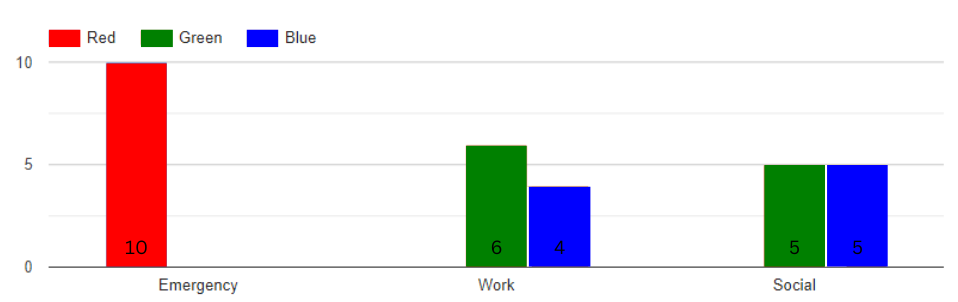}
    \caption{Most preferred color per category}
    \label{fig:prefer}
\end{figure}

 For both the Work and Emergency categories, participants indicated that the color Red conveyed a higher sense of importance for these categories. This finding is further supported by the follow-up question, where participants identified their most preferred color for each category, as illustrated in Figure \ref{fig:prefer}.



\begin{table}[h]
    \caption{5-point Likert scale of perceived importance}
    \label{tab:likert}
    \begin{tabular}{|l||l|l|l|l|l|}
        \hline
        \textbf{Category : Color} & \textbf{1} & \textbf{2} & 
            \textbf{3} & \textbf{4} & \textbf{5}\\ \hline \hline
        Emergency : Red & 0 & 0 & 1 & 0 & 9 \\ \hline
        Emergency : Green & 2 & 2 & 3 & 0 & 3 \\ \hline
        Emergency : Blue & 1 & 2 & 3 & 2 & 2 \\ \hline
        Work      : Red & 1 & 0 & 1 & 0 & 8 \\ \hline
        Work      : Green & 1 & 0 & 2 & 3 & 4 \\ \hline
        Work      : Blue & 1 & 4 & 2 & 1 & 2 \\ \hline
        Social    : Red & 2 & 1 & 2 & 2 & 3 \\ \hline
        Social    : Green & 2 & 1 & 2 & 2 & 3 \\ \hline
        Social    : Blue & 1 & 3 & 4 & 0 & 2 \\ \hline
    \end{tabular}
\end{table}

\begin{figure}[htbp]
    \centering
    \includegraphics[width=0.9\textwidth]{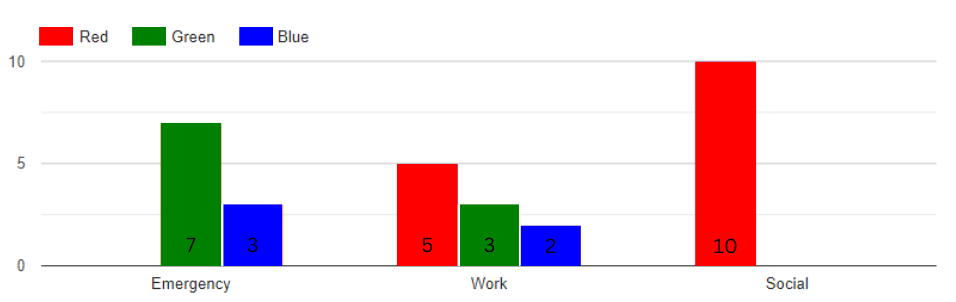}
    \caption{Least preferred color per category}
    \label{fig:least-prefer}
\end{figure}

\section {Design Implications and Limitations}

\par The conducted experiment provided valuable insights for improving notification system design. Color coding emerged as a key factor in enhancing user interaction, enabling users to quickly assess a notification's importance without needing to read it in detail. This is particularly beneficial for urgent alerts or during tasks requiring deep concentration~\cite{nasti2020discovering}. However, individual differences in color perception and associations may influence its effectiveness~\cite{yatid2012understanding}. Therefore, user customization of color schemes is recommended to personalize the experience and improve efficiency. Additionally, integrating AI to adapt notification delivery based on user behavior could optimize timing and presentation~\cite{kamal2021hybrid}, reducing disruptions while ensuring that essential information is delivered promptly.

\par Despite these insights, the study had several limitations. The sample consisted of a small, homogeneous group of university students, which limits the generalizability of the findings. A more diverse participant pool is needed to better understand how different demographics interact with notifications. The experiment's focus on a narrow set of notification categories and reliance on visual cues alone does not fully address the complexities of real-world notification management, which also involves varying types, priorities, and auditory or haptic feedback—elements particularly important for users with visual impairments. Furthermore, the controlled experimental setup may not accurately reflect real-life notification interactions and overlooks how users may adapt to notification cues over time, potentially affecting the long-term impact of design strategies on user attention and engagement.

\section{Conclusion and Recommendations}
\par In this experiment, we aimed to investigate how factors such as notification behavior and color influence user interaction and perception. Our goal was to determine whether these factors significantly impact how users interact with different notifications and if they should be incorporated into existing notification customization systems, such as the Focus Mode on Android and iOS.

\par The results of the experiment show that red notifications were more likely to be accepted than other colors, and notifications with non-urgent behavior were dismissed more frequently. Additionally, notifications with urgent behavior had faster average reaction times. Survey results revealed that all participants preferred the color red for emergency notifications, although it was less favored in other categories. Green and blue were equally preferred for work and social notifications, with the least preferred color for work notifications showing the most variability. Interestingly, we observed similar preferences for green and blue in both the experiment and the survey.

\par Based on these findings, we recommend incorporating the variables explored in this study as additional customization options for notifications, integrating them with existing features like Focus Mode. This would allow users to tailor notification behaviors and colors to their personal preferences, enhancing the overall notification experience.



\bibliography{main}




\end{document}